\pdfoutput=1
\documentclass[11pt,twoside,a4paper,cmspaper,final,collab]{cms-tdr}
\def\svnVersion{23ba5d7}\def\svnDate{2023/12/28}\def\cmsCernNoTag{CERN-EP-2023-031}\def\cmsCernDate{\today}\def\cmsMessage{Published in Physical Review Letters as \href{http://dx.doi.org/10.1103/PhysRevLett.131.262301}{\doi{10.1103/PhysRevLett.131.262301}.}}
\begin{document}\cmsNoteHeader{HIN-22-002}

\newlength\cmsTabSkip\setlength{\cmsTabSkip}{1ex}

\newcommand{\mumu}{\ensuremath{\PGmp\PGmm}\xspace}
\newcommand{\WgA} {\ensuremath{W^{\mathrm{A}}_{\PGg{}\mathrm{N}}}\xspace}
\newcommand{\WgPb} {\ensuremath{W^{\mathrm{Pb}}_{\PGg{}\mathrm{N}}}\xspace}
\newcommand{\RglPb} {\ensuremath{R^{\mathrm{Pb}}_{\Pg{}}}\xspace}
\newcommand{\PbPb}  {\ensuremath{\text{Pb-Pb}}\xspace}
\newcommand{\fD} {\ensuremath{f_{\mathrm{D}}}\xspace}
\newcommand{\fI} {\ensuremath{f_{\mathrm{I}}}\xspace}
\newcommand{\STARlight}{\textsc{STARlight}\xspace}

\title{Probing small Bjorken-\texorpdfstring{$x$}{x} nuclear gluonic structure via coherent \texorpdfstring{$\PJGy$}{J/psi} photoproduction in ultraperipheral Pb-Pb collisions at \texorpdfstring{$\sqrtsNN = 5.02\TeV$}{sqrt(sNN) = 5.02 TeV}}

\date{\today}

\abstract{Quasireal photons exchanged in relativistic heavy ion interactions are powerful probes of the gluonic structure of nuclei. 
The coherent $\PJGy$ photoproduction cross section in ultraperipheral lead-lead collisions is measured as a function of photon-nucleus center-of-mass energies per nucleon (\WgPb) over a wide range of $40<\WgPb<400\GeV$.
Results are obtained using data at the nucleon-nucleon center-of-mass energy of 5.02\TeV collected by the CMS experiment at the CERN LHC, corresponding to an integrated luminosity of $1.52\nbinv$. The cross section is observed to rise rapidly at low $\WgPb$, and plateau above $\WgPb \approx 40\GeV$, up to 400\GeV, entering a new regime of small Bjorken-$x$ (${\approx}6  \times 10^{-5}$) gluons being probed in a heavy nucleus. The observed energy dependence is not predicted by current quantum chromodynamic models.}

\hypersetup{%
pdfauthor={CMS Collaboration},%
pdftitle={Probing small Bjorken-x nuclear gluonic structure via coherent J/psi photoproduction in ultraperipheral Pb-Pb collisions at sqrt(s[NN]) = 5.02 TeV},%
pdfsubject={CMS},%
pdfkeywords={CMS, ultraperipheral, gluon saturation, nuclear shadowing, photon-nuclear}}

\maketitle 

The electromagnetic fields surrounding relativistic heavy ions are highly Lorentz contracted and can be effectively treated as 
linearly polarized quasireal photons~\cite{PhysRev.45.729, vonWeizsacker:1934nji}. 
These photons are powerful probes of the gluonic structure of nuclei~\cite{Klein:2020fmr}. 
At high energy colliders, when two ions pass each other at an impact parameter 
greater than twice the nuclear radius, it is referred to as an ``ultraperipheral collision'' or UPC. In UPCs, a photon from one ion can fluctuate into a virtual quark-antiquark pair and interact with the other ion via gluon exchanges before turning into a vector meson (VM) final state
~\cite{Adler:2002sc, Adamczyk:2017vfu, Afanasiev:2009hy, ALICE:2014eof, Sirunyan:2018sav, Sirunyan:2019nog, STAR:2022wfe, Abelev:2012ba, Abbas:2013oua, Acharya:2019vlb, Acharya:2019vlb,ALICE:2021gpt, Khachatryan:2016qhq, LHCb:2021bfl,LHCb:2022ahs}.
If the photon interacts with the whole nucleus, the process is known as a coherent interaction, 
while the incoherent process involves interactions of photons with the constituents of the target nucleus.
At energies accessible at the CERN LHC, studies of VM photoproduction in UPCs 
open the possibility of probing the poorly known nuclear gluonic structure and dynamics in the low Bjorken-$x$ region ($10^{-5}<x<10^{-2}$). 
For instance, the increasingly large gluon density at small $x$ may ultimately reach the unitarity limit, leading to saturation of the gluon density~\cite{Iancu:2003xm,Gelis:2010nm} and thus to saturation of the VM cross section. It has also been suggested that
the black-disk limit (BDL)~\cite{Gribov:1968gs,Frankfurt:2001nt,Frankfurt:2002wc,Frankfurt:2005mc,Rogers:2003vi,Alkin:2014rfa} of strong absorption may be reached at high energies, where the gluon density may not have saturated, but the scattering probability by multiple gluons becomes large enough that the cross section still saturates at the geometric limit of the nucleus. 

\begin{figure}[hbt!]
\centering
\includegraphics[width=\linewidth]{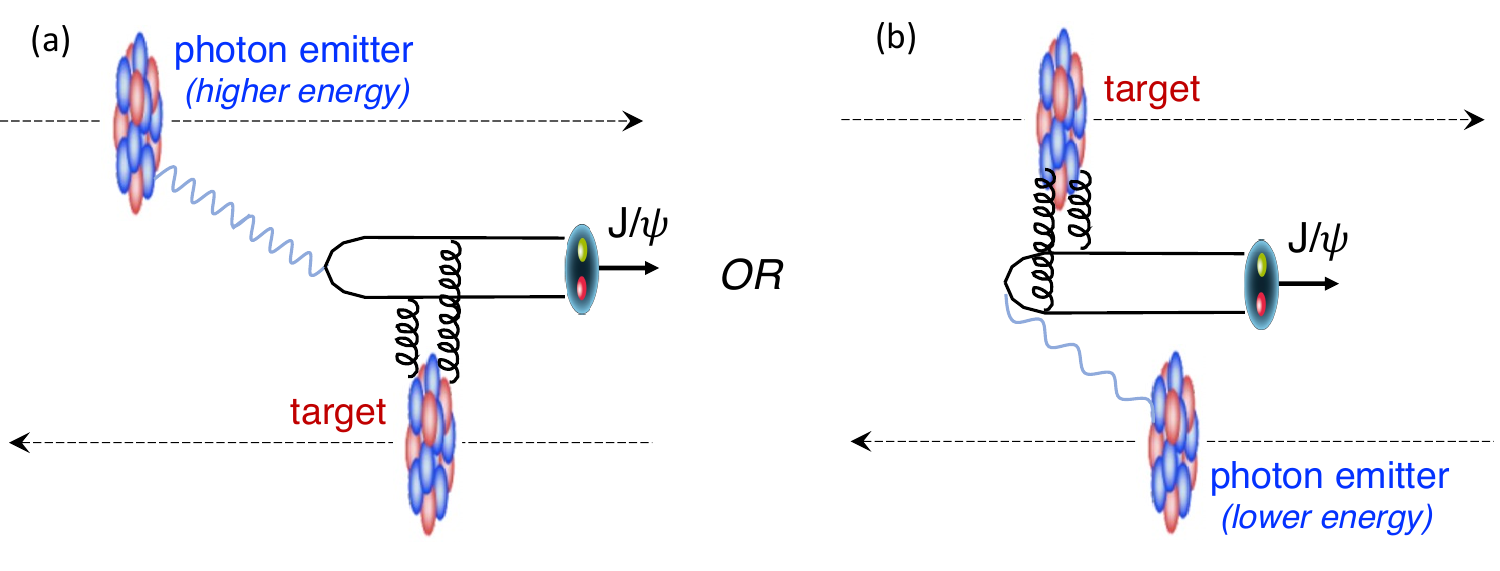}
\caption{A pictorial representation of the $\PJGy$ photoproduction process in Pb-Pb at leading-order QCD, showing the origin of the two-way ambiguity. The blue wavy and black helical lines represent emitted photons and gluons, respectively.\label{fig:cartoon}}
\end{figure}

The photoproduction of heavy-flavor VMs is of particular interest since it can be studied 
in the perturbative quantum chromodynamic (QCD) framework with a large energy scale defined by the quark mass.
In the leading-order QCD contribution, which is sketched in Fig.~\ref{fig:cartoon}, the coherent $\PJGy$ photoproduction cross section is proportional to the square of the gluon density function.
In photon-proton interactions, the $\PJGy$ production cross section significantly increases with the photon-proton center-of-mass energy~\cite{ZEUS:2002wfj, H1:2005dtp, H1:2013okq, LHCb:2013nqs, ALICE:2014eof, ALICE:2018oyo}. This results from
the rapidly rising gluon density in the proton at small $x$~\cite{Bjorken:1968dy,Frankfurt:2005mc}.
For photons scattering off heavy-nuclear targets, while significant nuclear suppression effects have been observed for coherent $\PJGy$ photoproduction in UPCs by the ALICE~\cite{Abbas:2013oua,Abelev:2012ba,Acharya:2019vlb,ALICE:2021gpt}, CMS~\cite{Khachatryan:2016qhq}, and LHCb~\cite{LHCb:2021bfl,LHCb:2022ahs} experiments, the underlying mechanism remains
poorly understood. Little is known regarding the behavior of the interaction toward the high energy limit, which probes gluons at small $x$ (i.e., $x<10^{-4}$). 
The major limitation is the experimental difficulty in distinguishing the photon-emitter nucleus and target nucleus in a symmetric collision system~\cite{Guzey:2013jaa}. 
For VM photoproduction off nuclei at a given rapidity, there is a mixture of 
low and high energy photon contributions, as shown by the opposing sides of Fig.~\ref{fig:cartoon}.
This two-way ambiguity limits measurements at LHC energies to below 125\GeV 
in the photon-nucleus center-of-mass energy per nucleon (\WgA).

A novel approach for resolving this two-way ambiguity was proposed in Refs.~\cite{Guzey:2013jaa,Guzey:2016piu}.
By controlling the UPC impact parameter using forward-emitted neutron multiplicities from the ions' electromagnetic dissociation (EMD),
the relative contributions of low and high energy photons to the measured $\PJGy$ cross section can be separated without ambiguity.

The results in this Letter use Pb-Pb UPCs at a nucleon-nucleon center-of-mass energy $\sqrtsNN = 5.02\TeV$, which were collected by the CMS experiment at the LHC in 2018. The data correspond to an integrated luminosity of $1.52\nbinv$.
By applying the aforementioned forward neutron tagging technique, this Letter reports the cross section measurement of the coherent $\PJGy$ photoproduction off Pb nuclei over a broad energy range of $40<\WgPb<400\GeV$, which includes previously unexplored small-$x$ regions (i.e., $x<10^{-4}$). Tabulated results are provided in the HEPData record for this analysis~\cite{hepdata}.

The CMS apparatus~\cite{CMS:2008xjf} is a multipurpose, nearly hermetic detector, designed to trigger on~\cite{CMS:2020cmk,Khachatryan:2016bia} and identify electrons, muons, photons, and hadrons~\cite{CMS:2020uim,CMS:2018rym,CMS:2014pgm}. A global ``particle-flow" algorithm~\cite{CMS:2017yfk} aims to reconstruct all individual particles in an event, combining information provided by the all-silicon inner tracker and by the crystal electromagnetic and brass-scintillator hadron calorimeters, operating inside a 3.8\unit{T} superconducting solenoid, with data from the gas-ionization muon detectors embedded in the flux-return yoke outside the solenoid. 
Forward calorimeters~\cite{Bayatian:2006jz}, made of steel and quartz fibers, extend the pseudorapidity coverage provided by the barrel and end-cap detectors.
Two zero-degree calorimeters (ZDCs)~\cite{Suranyi:2021ssd}, made of quartz fibers and plates embedded in tungsten absorbers, are used to detect neutrons from nuclear dissociation events. 

Events are selected using a hardware-based trigger system that requires at least one muon candidate with no explicit selection on its transverse momentum (\pt), coincident with a \PbPb bunch crossing~\cite{Khachatryan:2016bia}. At the trigger level, events with an energy deposit above the noise threshold in either of the forward calorimeters are vetoed. 
For the off-line analysis, events must have a primary interaction vertex, formed by using two or more tracks coming from the collision, that is within 20\unit{cm} along the beam axis and at a radius less than 2\unit{cm} in the transverse plane from the detector center.
To suppress hadronic collisions, the largest energy deposits in the forward calorimeters must be below 7.3 and 7.6\GeV in the positive and negative rapidity sides, respectively. These energy requirements are determined from empty bunch-crossing events~\cite{CMS:2020skx}, providing a purity of ${\approx}100\%$ with negligible efficiency loss. Finally, events must contain exactly two muon candidates and no additional tracks in the range $\abs{\eta}<2.4$.

To select high-quality muons, both muons must be ``soft muons," as defined in Ref.~\cite{Chatrchyan:2012xi}, 
based on the combined information of the tracker and muon detectors.
Pairs of muons of opposite signs with an invariant mass in the range $2.6<m_{\mumu}<4.2\GeV$ are taken as $\PJGy$ and $\PGyP{2S}$ meson candidates.
To reconstruct $\PJGy$ mesons at low \pt, the dimuon rapidity must satisfy $1.6<\abs{y}<2.4$ for both muons to reach the muon detectors.
For each muon pair, at least one of the muon candidates must match the triggered muon. 
Combinatorial backgrounds are estimated using same-sign muon pairs and are negligible after the event and muon requirements. 

The raw number of $\PJGy$ signal events contains contributions from several physics processes: 
coherent and incoherent (with or without nucleon dissociation) $\PJGy$ production from direct photon-nucleus interactions, and also $\PJGy$ mesons coming from the decay of $\PGyP{2S}$ mesons, referred to as ``feed-down.'' The coherent $\PJGy$ meson raw yields ($N^{\text{coh}}_{\PJGy}$) are extracted by fitting the reconstructed dimuon invariant mass and \pt distributions for each given rapidity bin, as done in Refs.~\cite{Acharya:2019vlb,ALICE:2021gpt} (see Appendix for details).
The detector reconstruction efficiency is estimated using a Monte Carlo sample of coherent $\PJGy \to \PGmp\PGmm$ events generated with the \STARlight (v3.13) event generator~\cite{Klein:2016yzr}. These simulated events are further processed with the full CMS detector response simulation using \GEANTfour~\cite{Agostinelli:2002hh}.

The coherent $\PJGy$ production differential cross section is determined using
\begin{equation}
\frac{\rd \sigma^{\text{coh}}_{\PJGy}}{\rd y} = \frac{N^{\text{coh}}_{\PJGy}}{\epsilon_{\PJGy} \: \mathcal{B}_{\PJGy \to \PGm\PGm} \: \lumi_{\text{int}} \: \epsilon_\text{evtsel} \: \Delta y},
\end{equation}
where the event selection efficiency of $\epsilon_\text{evtsel} = 94.2\%$ and the $\PJGy$ reconstruction efficiency $\epsilon_{\PJGy}$, ranging from $2.6\%$ to $11.9\%$ for $1.6<\abs{y}<2.4$, are evaluated using simulated samples. 
Additional corrections determined using the tag-and-probe technique~\cite{CMS:2011aa,Chatrchyan:2012xi} are applied to account for any data-to-simulation discrepancy.
The world-average value for the \PJGy dimuon branching fraction~\cite{PDG2022} $\mathcal{B}_{\PJGy \to \mumu}$ is used.
The quantity $\lumi_{\text{int}}$ is the total integrated luminosity of the data sample~\cite{CMS:2022bjp}, and $\Delta y$ is the rapidity binning used in the measurement.

Coherent $\PJGy$ photoproduction cross sections are reported in classes of the forward-neutron multiplicity, 
which is determined with ZDC energy deposits, as done in Ref.~\cite{CMS:2020skx}. 
Based on the neutron peaks observed in the total ZDC energy distribution, events are classified as having no neutrons (0n) or with at least one neutron (Xn, X $\geq 1$) in each ZDC. 
The purities of 0n and Xn classes are estimated using a multi-Gaussian 
function fit to the ZDC energy distribution, and are ${>}99.6\%$~\cite{CMS:2020skx}.
Three neutron multiplicity event classes are formed by combining the classifications of each ZDC side, labeled as 0n0n, 0nXn, and XnXn. 
Results without neutron multiplicity selection, labeled as AnAn, are also reported for comparison and complementarity with previous measurements.
The large EMD cross section ($\sim$200\unit{b})~\cite{Pshenichnov:2001qd,ALICE:2012aa,ATLAS:2020epq} leads to concurrent neutron emission from other \PbPb interactions in the same bunch crossing, which can migrate the neutron multiplicity class to a higher one. Such EMD pileup effects can be corrected following the approach employed in Ref.~\cite{CMS:2020skx} using the zero-bias triggered data (requiring only that two beams cross each other), where the migration probability is directly determined by the probability of observing each neutron multiplicity class. 
The relative fractions of $\PJGy$ candidates in 0n0n, 0nXn, and XnXn classes are determined to be 73, 21, and 6\% corrected from the corresponding raw fractions of 65, 28, and 7\%, respectively.

Because of the symmetry of the Pb-Pb collision system, a $\PJGy$ meson measured at rapidity $y$ can result from two possible photon energies: $\omega_{1} = (M_{\PJGy}/2)\exp(-y)$ and $\omega_{2} = (M_{\PJGy}/2)\exp(+y)$. 
The $x$ of the associated parton and $\WgPb$ with a photon of energy $\omega_1$ can be found using $x=(M_{\PJGy}/\sqrtsNN)\exp(+y)$ and $\WgPb = \sqrt{\smash[b]{\sqrtsNN M_{\PJGy} \exp(-y)}}$. The corresponding values for a photon of energy $\omega_2$ are obtained by changing the sign of $y$ in each expression. 

The measured differential cross section at a rapidity $y$ for a neutron multiplicity class, $i\textrm{n}j\textrm{n}$ ($=$ 0n0n, 0nXn, XnXn), of UPC events is the sum of both photon energy contributions~\cite{Guzey:2013jaa,Guzey:2016piu}:
\begin{equation}
\label{eq:xsec_inclusive_y}
    \frac{\rd \sigma_{\PJGy}^{i\textrm{n}j\textrm{n}}(y)}{\rd y} =  n^{i\textrm{n}j\textrm{n}}_{\PGg{}\mathrm{A}}(\omega_{1}) \: \sigma_{\PJGy}(\omega_{1}) 
	+ n^{i\mathrm{n}j\mathrm{n}}_{\PGg{}\mathrm{A}}(\omega_{2})  \: \sigma_{\PJGy}(\omega_{2}), 
\end{equation}
where $\omega_{1}$ and $\omega_{2}$ are themselves functions of $y$. Here, $n_{\PGg{}\mathrm{A}}(\omega) = \omega \ddinline{N_{\PGg{}\mathrm{A}}(\omega)}{\omega}$ is the photon flux and $\sigma_{\PJGy}(\omega)$ is the coherent $\PJGy$ photoproduction cross section for a single $\PGg$-ion interaction, averaged over the given $y$ range.
The photon flux from a relativistic ion can be calculated using \STARlight~\cite{Klein:2016yzr}, which implements an equivalent approximation model~\cite{Cahn:1990jk,Baur:1990fx,Vidovic:1992ik,Baltz:2002pp}.
The flux strongly depends on the ion-ion impact parameter; the fraction of high energy photon flux increases at smaller impact parameters and for higher neutron multiplicities.
Therefore, at each given rapidity, after measuring the $\PJGy$ differential coherent cross section (i.e., the left-hand side of Eq.~\eqref{eq:xsec_inclusive_y}) and computing the corresponding photon fluxes in each of the three neutron classes, a combined $\chi^2$ fit of $\rd \sigma_{\PJGy}^{i\textrm{n}j\textrm{n}}(y) / \rd y$ versus $n^{i\textrm{n}j\textrm{n}}_{\PGg{}\mathrm{A}}(\omega_{1})$ and $n^{i\textrm{n}j\textrm{n}}_{\PGg{}\mathrm{A}}(\omega_{2})$ is performed to extract the differential cross sections for low and high energies, $\sigma_{\PJGy}(\omega_{1})$ and $\sigma_{\PJGy}(\omega_{2})$.
The statistical uncertainty is evaluated from the combined $\chi^2$ fit, and depending on the $y$ range, amounts to 2\%--4\% and 6\%--9\% for $\sigma_{\PJGy}(\omega_{1})$ and $\sigma_{\PJGy}(\omega_{2})$, respectively.

Systematic uncertainties are evaluated by taking the maximum variation from the nominal result for a given source.
The uncertainties are 1.5\% in the integrated luminosity~\cite{CMS:2021xjt,CMS:2022bjp}, and 0.55\% in $\mathcal{B}_{\PJGy \to \mumu}$~\cite{PDG2022}.
Uncertainties in the tag-and-probe corrections are 3\%--8\% from low to high \WgPb  values.
The choice of the forward calorimeter threshold energy used to reject hadronic contamination leads to a 1\%--4\% uncertainty.
For the systematic uncertainty in the raw signal yield, different functional forms for modeling the shapes of the signal and background are studied, including: 
(i) a sum of Crystal Ball~\cite{Oreglia:1981fx} and Gaussian functions for the $\PJGy$ signal shape; 
(ii) a second-order polynomial for the dimuon continuum background;
(iii) fixing the Crystal Ball shape parameters to the values determined from Monte Carlo signal distributions; 
(iv) varying the range of the dimuon invariant mass distribution used in the fit, 
and (v) changing the dimuon continuum \pt shape to the
one determined from sideband mass regions. 
The results vary by less than 8\%.
The uncertainty in the neutron migration correction is estimated to be ${<}2\%$ by comparing to an alternative approach
following Ref.~\cite{ATLAS:2020epq}, using the observed rate of single- and double-sided neutron emissions per bunch crossing.
Individual sources of experimental uncertainties are added in quadrature to obtain the total systematic uncertainty of 6\%--12\%.
Theoretical uncertainties in the cross section measurement from the photon flux calculation are separately evaluated to be 3\%--16\% by varying the Pb nuclei radius ($R=6.67\pm0.03$\unit{fm}), the nuclear skin thickness ($a=0.56\pm0.03$\unit{fm})~\cite{dEnterria:2020dwq}, and the EMD cross section~\cite{Baltz:1996as} within the experimental uncertainties, where uncertainties in the extracted cross sections at low and high \WgPb  are found to be strongly anticorrelated.
Correlations between uncertainties are evaluated as functions of $y$ for each neutron multiplicity class or \WgPb (which also captures correlations between neutron multiplicity classes) and included in the covariance matrix~\cite{hepdata} of the fit.

\begin{figure*}[hbt!]
\centering
\includegraphics[width=0.49\textwidth]{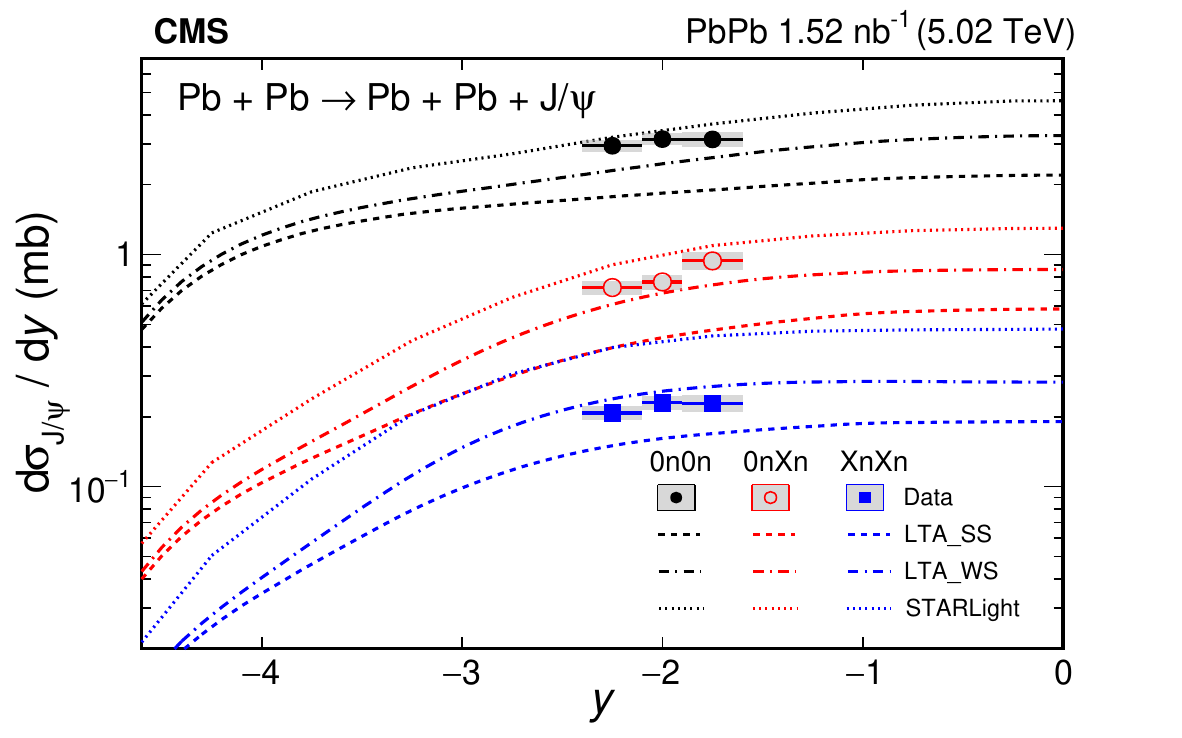}
\includegraphics[width=0.49\textwidth]{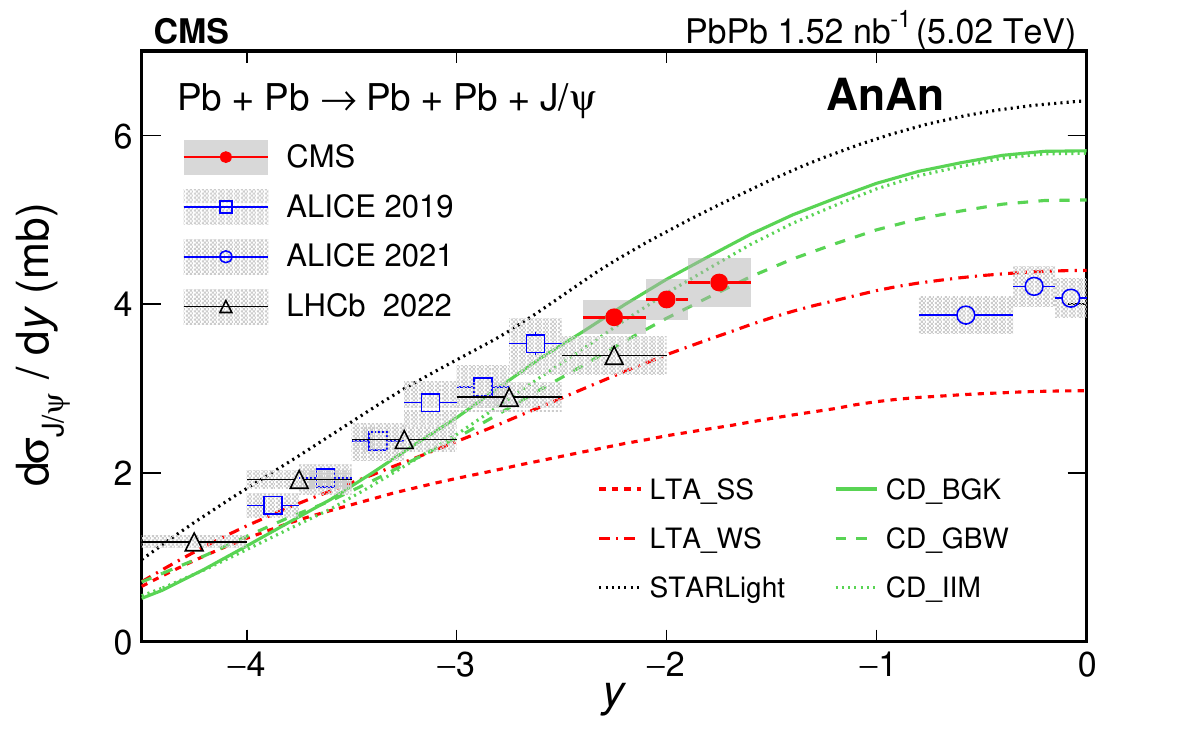}
\caption{The differential coherent $\PJGy$ photoproduction cross section as a function of rapidity, in different neutron multiplicity classes: 0n0n, 0nXn, and XnXn (left); AnAn (right). The small vertical bars and shaded boxes represent the statistical and systematic uncertainties, respectively.
The horizontal bars show the bin widths.
Theoretical predictions from LTA weak and strong shadowing~\cite{Guzey:2016piu}, color dipole models (CD\_BGK,CD\_BGW, and CD\_IIM)~\cite{Luszczak:2019vdc}, and \STARlight~\cite{Klein:2016yzr} are shown by the curves. The right plot also displays the results from the ALICE~\cite{Acharya:2019vlb,ALICE:2021gpt} and LHCb~\cite{LHCb:2022ahs} experiments.\label{fig:dSigdY}}
\end{figure*}

The measured coherent $\PJGy$ photoproduction differential cross sections over the rapidity range $1.6<\abs{y}<2.4$ are reported in Fig.~\ref{fig:dSigdY} (left) in three neutron multiplicity classes. The results with no neutron selection are given in the right plot of Fig.~\ref{fig:dSigdY}.
The measurements are performed as a function of $\abs{y}$, but are shown at negative $y$ values, following the ALICE convention~\cite{Acharya:2019vlb,ALICE:2021gpt}. 
Predictions from several theoretical calculations are also shown in both plots for comparison. 
The leading twist approximation (LTA)~\cite{Guzey:2016piu} 
is a perturbative QCD calculation that takes 
into account nuclear shadowing effects from multinucleon interference. 
Both weak (WS) and strong (SS) shadowing scenarios are shown~\cite{Guzey:2016piu}.
The color dipole (CD) models, with different model parameters (BGK, BGW, IIM), assume quark-antiquark dipole scattering from the nuclear targets~\cite{Luszczak:2019vdc}.
 
For the case of no neutron selection (AnAn), the data follow the trend of the forward-rapidity measurements from ALICE~\cite{Acharya:2019vlb} over a new $y$ region. 
None of the models describe the combined results over the full rapidity range.
The color dipole models agree with the measurements in the forward-rapidity region, but fail to describe the data at $y \approx 0$. 
In each neutron multiplicity class, the LTA predictions tend to be lower than the CMS results, particularly for the strong shadowing scenario. 
These comparisons indicate that there are key ingredients missing from the theoretical understanding of high energy photon-nucleus scattering processes.

\begin{figure}[hbt!]
	\centering
	\includegraphics[width=\linewidth]{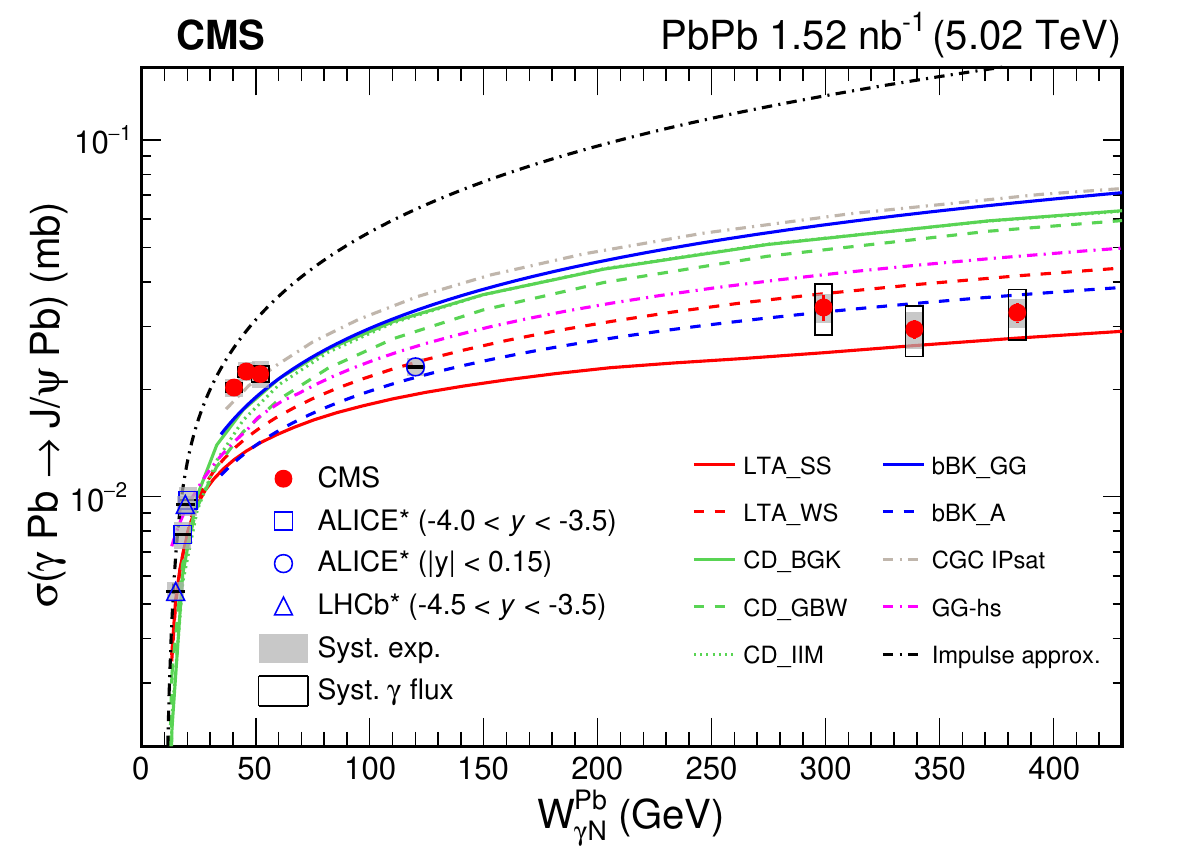}
	\caption{The total coherent $\PJGy$ photoproduction cross section as a function of \WgPb from 
	the CMS measurement in Pb-Pb UPCs at $\sqrtsNN = 5.02\TeV$.
    Approximated results (implied by the asterisk) from the ALICE~\cite{Acharya:2019vlb,ALICE:2021gpt} and LHCb~\cite{LHCb:2022ahs} experiments are displayed for specific rapidity regions, where the two-way ambiguity effect is expected to be negligible.
	The \WgPb values used correspond to the center of each experiment's rapidity range. The vertical bars and the shaded and open boxes represent the statistical, experimental, and theoretical (photon flux) uncertainties, respectively. The predictions from various theoretical calculations~\cite{Guzey:2013jaa,Mantysaari:2017dwh,Cepila:2017nef,Guzey:2016piu,Bendova:2020hbb,Luszczak:2019vdc} are shown by the curves.}
	\label{fig:xsec_vs_W}
\end{figure}

To gain further insight, the total measured $\PJGy$ coherent photoproduction cross section as a function of $\WgPb$ up to $\approx$400\GeV is shown in Fig.~\ref{fig:xsec_vs_W}, after decomposing the two-way ambiguity.
Because the contributions of high energy photons are negligible at very forward rapidity 
(less than 5\% for $-4.5<y<-3.5$)~\cite{Guzey:2013jaa,Guzey:2016piu}, and the fact that at $y \approx 0$, $\omega_{1} \approx \omega_{2} \approx M_{\PJGy}/2$,
the total cross section at lower \WgPb values can be approximated using ALICE and LHCb measurements. These results are also shown in Fig.~\ref{fig:xsec_vs_W}. 
The experimental and theoretical (from the photon flux) uncertainties are displayed separately in Fig.~\ref{fig:xsec_vs_W}.
Predictions from the LTA and CD models, as well as the gluon saturation models bBK~\cite{Bendova:2020hbb}, IPsat~\cite{Mantysaari:2017dwh}, and GG~\cite{Cepila:2017nef}), are compared to the experimental measurements. 
The prediction ($\sigma^{\text{IA}}$) from the impulse approximation (IA) model~\cite{Guzey:2013jaa} is also shown, 
based on a simple scaling of the experimental data from exclusive \PJGy photoproduction off protons 
with the nuclear form factor and neglecting all other nuclear effects, except for coherence.

The measured total cross section has an unexpected energy dependence, approximately quadrupling as \WgPb goes from 15 to 40\GeV. 
This is consistent with the expectation of a fast-growing gluon density at low $x$ (\eg, from the IA model). 
However, this trend vanishes for $\WgPb>40\GeV$, and instead the total cross section begins a slow linear rise with a slope of $(2.2 \pm 1.9) \times 10^{-5}$\unit{mb}/\GeV determined by a fit to CMS data with proper consideration of the covariance matrix of both statistical and systematic uncertainties~\cite{hepdata}. Considering the experimental uncertainties across the measured $\WgPb$ range, none of the theoretical models are consistent with the measurements, with the CD-BGK model having the best $p$ value of $1.6  \times 10^{-8}$. 
This could imply the onset of novel physics mechanisms in the coherent $\PJGy$ photoproduction process starting at $\WgPb \approx 40\GeV$, for example, the saturation of the gluon density in the Pb nucleus at the corresponding $x$ value.
This picture may be less straightforward due to higher-order perturbative QCD corrections (\eg, quark-antiquark exchange)~\cite{Eskola:2022vpi,Eskola:2022vaf}. On the other hand, the observed behavior is expected when approaching the BDL, where a major part of the target nucleus becomes completely absorptive to photons because of the large scattering probability in the presence of dense gluons at small $x$~\cite{Frankfurt:2001nt}.
In this scenario, the photon-nucleus cross section approaches the unitarity limit allowed by the geometric size of the nucleus. 
The slow rise in the measured cross section from about 40 to 400\GeV would then be due to the periphery of the nucleus not becoming fully ``black.''

\begin{figure}[hbt!]
	\centering
	\includegraphics[width=\linewidth]{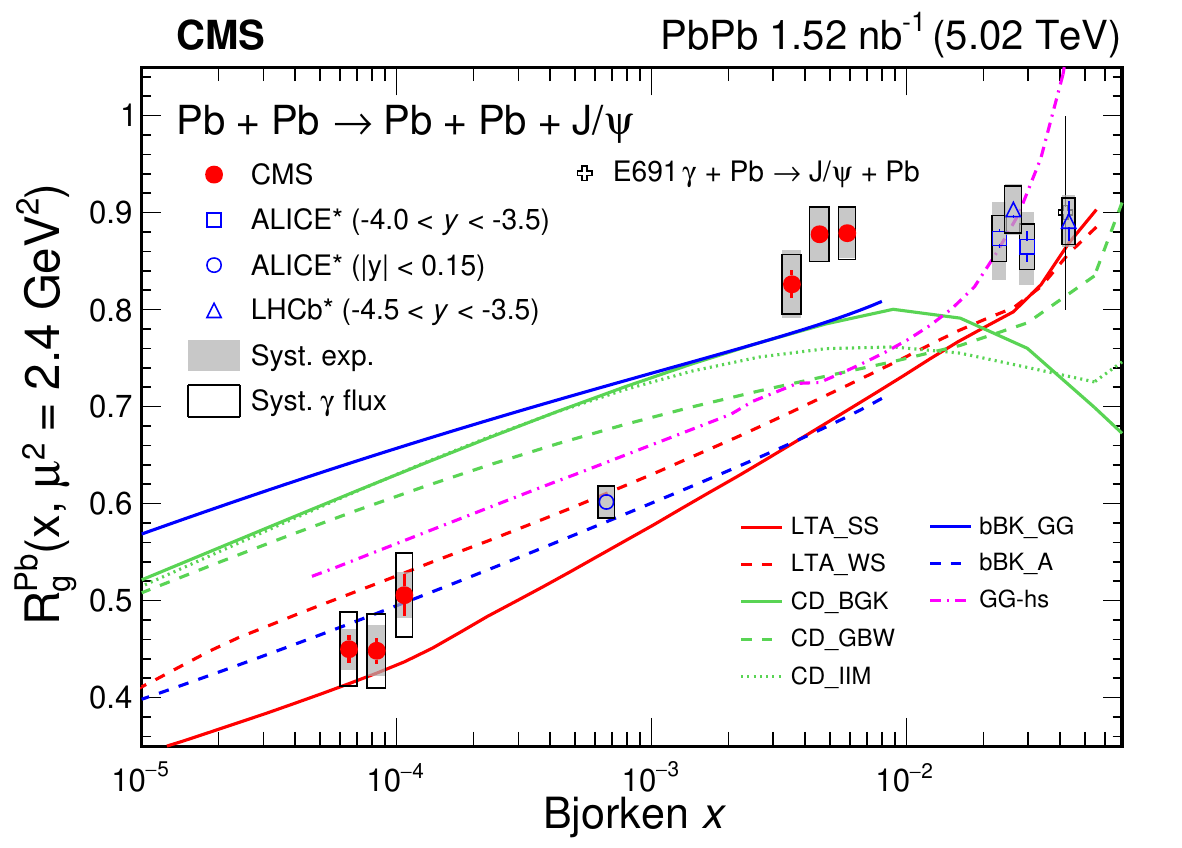}
	\caption{The nuclear gluon suppression factor $\RglPb$ as a function of Bjorken $x$ extracted from the CMS measurement of the coherent $\PJGy$ photoproduction in Pb-Pb UPCs at $\sqrtsNN=5.02$\TeV.
    Approximated results (implied by the asterisk) from the ALICE~\cite{Acharya:2019vlb,ALICE:2021gpt} and LHCb~\cite{LHCb:2022ahs} experiments are displayed for specific rapidity regions, where the two-way ambiguity effect is expected to be negligible.
	A measurement of $\PJGy$ photoproduction from the E691 experiment~\cite{Guzey:2020ntc,FermilabTaggedPhotonSpectrometer:1986xzf} is also given.
	The $x$ values are evaluated at the centers of their corresponding rapidity ranges.
	The vertical bars and shaded and open boxes represent the statistical, experimental systematic, and theoretical systematic uncertainties, respectively.
	The latter is due to the uncertainties in the photon flux and the IA. 
	Predictions from various theoretical calculations~\cite{Cepila:2017nef,Guzey:2016piu,Bendova:2020hbb,Luszczak:2019vdc} are shown by the curves.}
	\label{fig:suppression_vs_x}
	\end{figure}

To quantify the nuclear effects on the observed gluon density function in a Pb nucleus, a nuclear gluon suppression factor, $\RglPb(x, \PGm^{2} = 2.4 \GeV^{2})$ as a function of $x$ and with $\PGm=M_{\PJGy}/2$, where $\PGm$ is the energy scale, is defined as $\RglPb = \sqrt{\smash[b]{\sigma^{\text{Meas}}/\sigma^{\text{IA}}}}$ in Ref.~\cite{Guzey:2013xba}, where $\sigma^{\text{Meas}}$ is the measured cross section.
The extracted $\RglPb$ values are shown in Fig.~\ref{fig:suppression_vs_x} as a function of $x$.
The suppression in the relatively high-$x$ (low $\WgPb$) region of $x>5 \times 10^{-3}$ is approximately 0.8--0.9.
At smaller $x$ values, $\RglPb$ starts dropping rapidly to 0.4--0.5 for $x \approx 6  \times 10^{-5}$. Similar to the cross section measurement, no theoretical model 
predicts the observed values and $x$ dependence of $\RglPb$ over the wide $x$ range reported.
If the BDL of strong absorption is reached for $x<5 \times 10^{-3}$, corresponding to $\WgPb>40\GeV$, the direct connection between the measured cross section and the single gluon density function, as in the weak absorption limit, no longer holds.  Instead, the cross section will reach the unitarity limit while the gluon density can continue increasing. Therefore, new theoretical approaches are needed to understand the structure and dynamics of the strong force for a heavy nucleus in this new domain of extreme gluon density.

In summary, the measurement of the coherent $\PJGy$ photoproduction cross section off lead nuclei, as a function of the photon-nuclear center-of-mass energy per nucleon ($\WgPb$) has been presented over a broad energy range. In a coherent process, the $\PJGy$ is produced by the photon interacting with the whole nucleus.
Results are obtained with ultraperipheral Pb-Pb collision data by applying the forward neutron tagging technique. 
The cross section is observed to rise rapidly at low $\WgPb$ but appears to plateau above $\WgPb \approx 40\GeV$, up to 400\GeV, where a new regime of gluon momentum fraction (Bjorken $x \approx  6 \times 10^{-5}$) in a heavy nucleus is probed.
This observed trend is not predicted by current theoretical models. 
This can be interpreted either as the first direct evidence for the saturation of nuclear gluonic density or that the scattering cross section is near the black-disk limit.

While this analysis was under journal review, the ALICE Collaboration published a related analysis~\cite{ALICE:2023jgu} and their results are consistent with our observed data trend.

\begin{acknowledgments}
	We congratulate our colleagues in the CERN accelerator departments for the excellent performance of the LHC and thank the technical and administrative staffs at CERN and at other CMS institutes for their contributions to the success of the CMS effort. In addition, we gratefully acknowledge the computing centers and personnel of the Worldwide LHC Computing Grid and other centers for delivering so effectively the computing infrastructure essential to our analyses. Finally, we acknowledge the enduring support for the construction and operation of the LHC, the CMS detector, and the supporting computing infrastructure provided by the following funding agencies: BMBWF and FWF (Austria); FNRS and FWO (Belgium); CNPq, CAPES, FAPERJ, FAPERGS, and FAPESP (Brazil); MES and BNSF (Bulgaria); CERN; CAS, MoST, and NSFC (China); MINCIENCIAS (Colombia); MSES and CSF (Croatia); RIF (Cyprus); SENESCYT (Ecuador); MoER, ERC PUT and ERDF (Estonia); Academy of Finland, MEC, and HIP (Finland); CEA and CNRS/IN2P3 (France); BMBF, DFG, and HGF (Germany); GSRI (Greece); NKFIH (Hungary); DAE and DST (India); IPM (Iran); SFI (Ireland); INFN (Italy); MSIP and NRF (Republic of Korea); MES (Latvia); LAS (Lithuania); MOE and UM (Malaysia); BUAP, CINVESTAV, CONACYT, LNS, SEP, and UASLP-FAI (Mexico); MOS (Montenegro); MBIE (New Zealand); PAEC (Pakistan); MES and NSC (Poland); FCT (Portugal);  MESTD (Serbia); MCIN/AEI and PCTI (Spain); MOSTR (Sri Lanka); Swiss Funding Agencies (Switzerland); MST (Taipei); MHESI and NSTDA (Thailand); TUBITAK and TENMAK (Turkey); NASU (Ukraine); STFC (United Kingdom); DOE and NSF (USA). Individuals were supported by the National Science Center (Poland), grants no. 2015/19/N/ST2/02697 and 2018/28/T/ST2/00199.
\end{acknowledgments}
\bibliography{auto_generated}

\appendix
\section{Method of coherent \texorpdfstring{\PJGy}{J/psi} yield extraction \label{appendix:a}}

To extract the $\PJGy$ yield from coherent photoproduction, a two-step analysis is performed by fitting first the dimuon invariant mass distribution at $\pt<0.2\GeV$, and then the dimuon \pt distribution within the mass window of $2.95<m_{\mumu}<3.25\GeV$. The raw signal yields of $\PJGy$ and $\PGyP{2S}$ mesons, $N_{\PJGy}$ and $N_{\PGyP{2S}}$, are extracted by fitting the dimuon invariant mass distribution for a given rapidity bin with two Crystal Ball functions to describe the $\PJGy$ and \PGyP{2S} resonances, and a third-order polynomial function to parametrize the underlying background from quantum electrodynamic processes ($\PGg\PGg \to \mumu$)

\begin{figure*}[hbt!]
	\centering
	\includegraphics[width=0.49\textwidth]{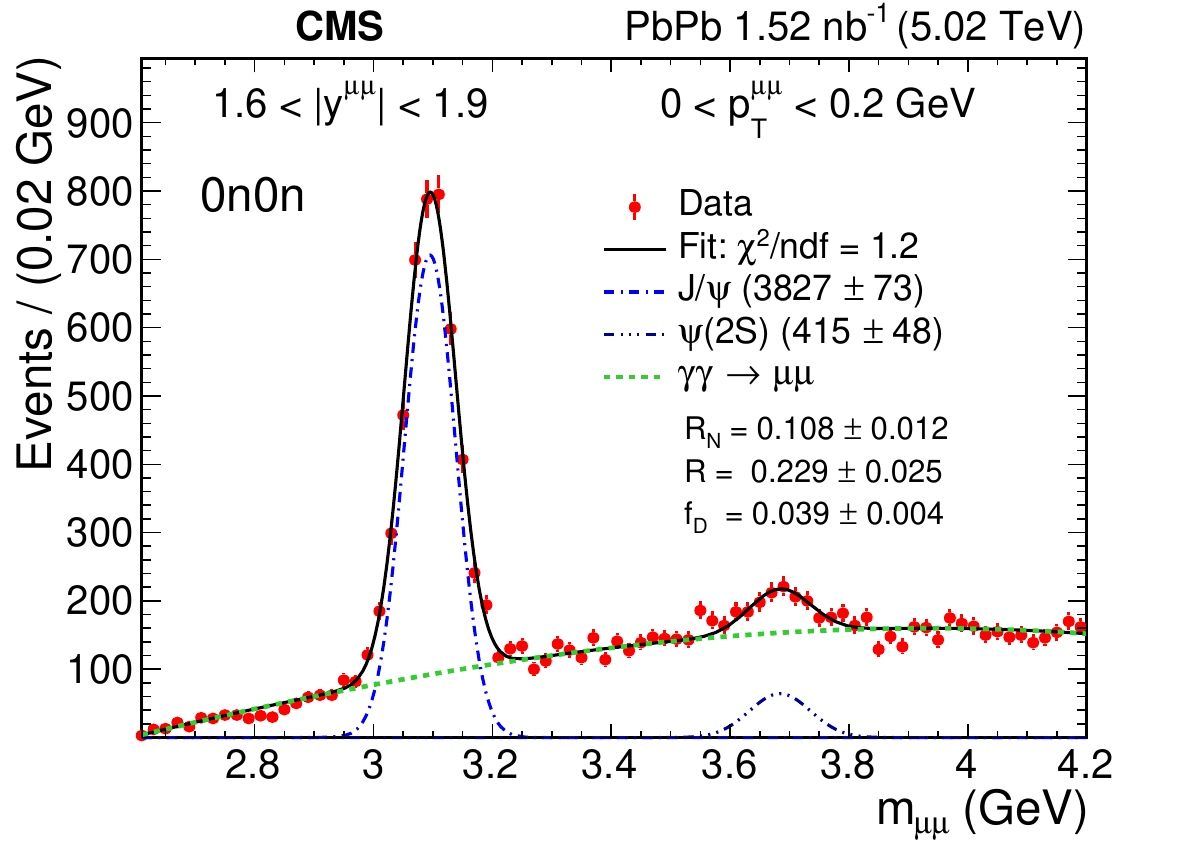}
	\includegraphics[width=0.49\textwidth]{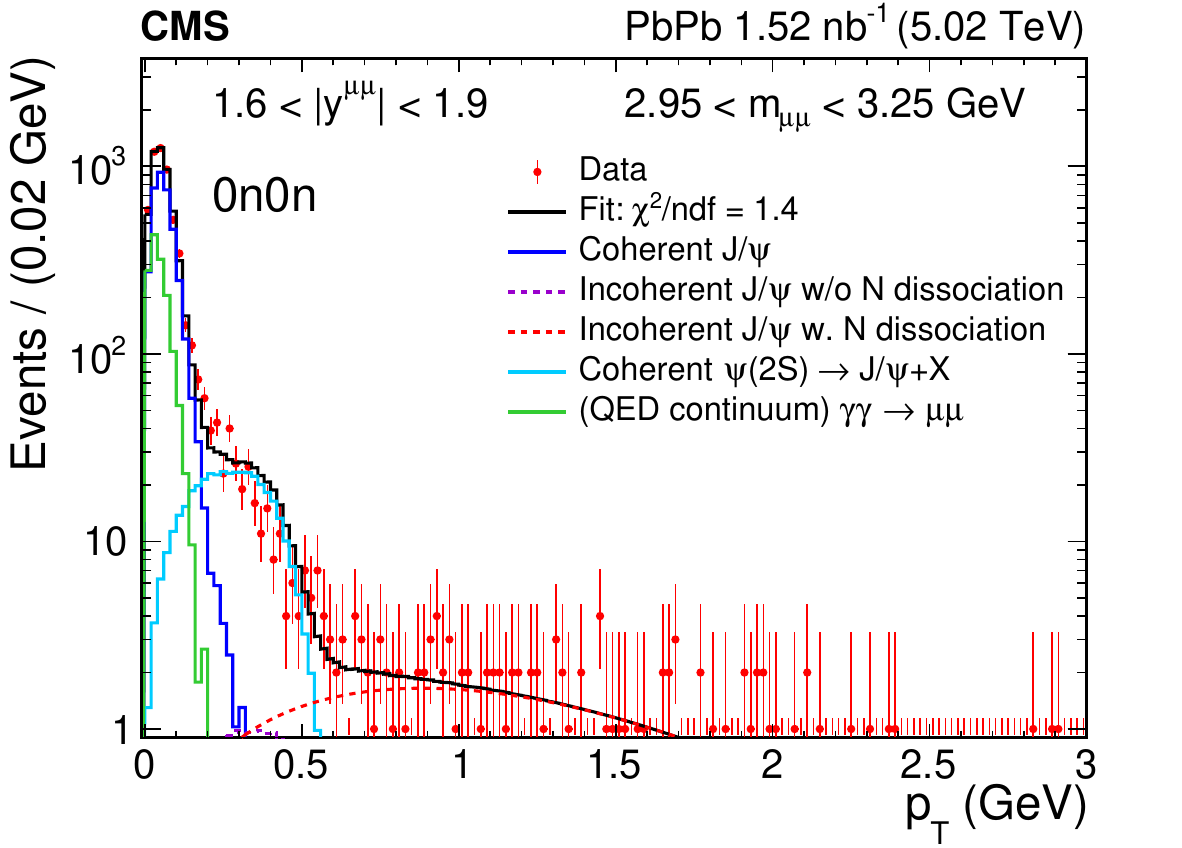}
	\includegraphics[width=0.49\textwidth]{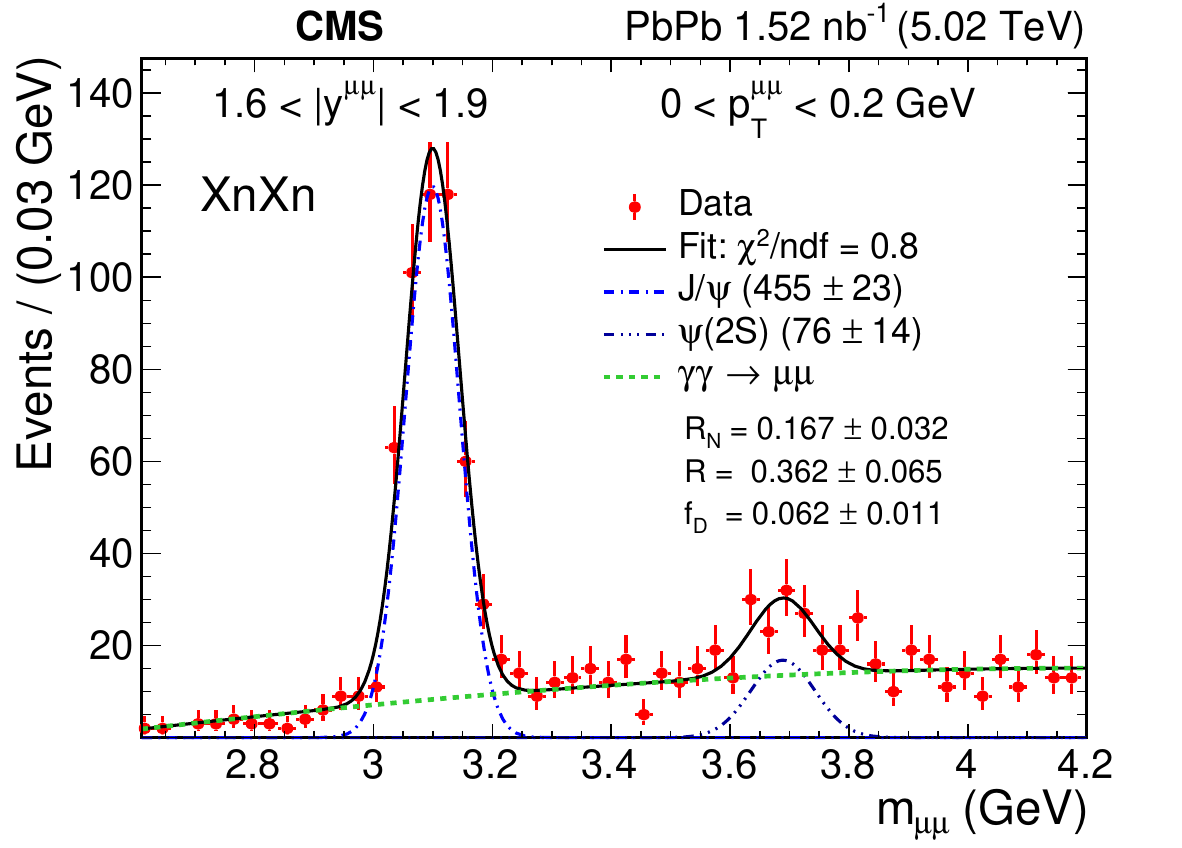}
	\includegraphics[width=0.49\textwidth]{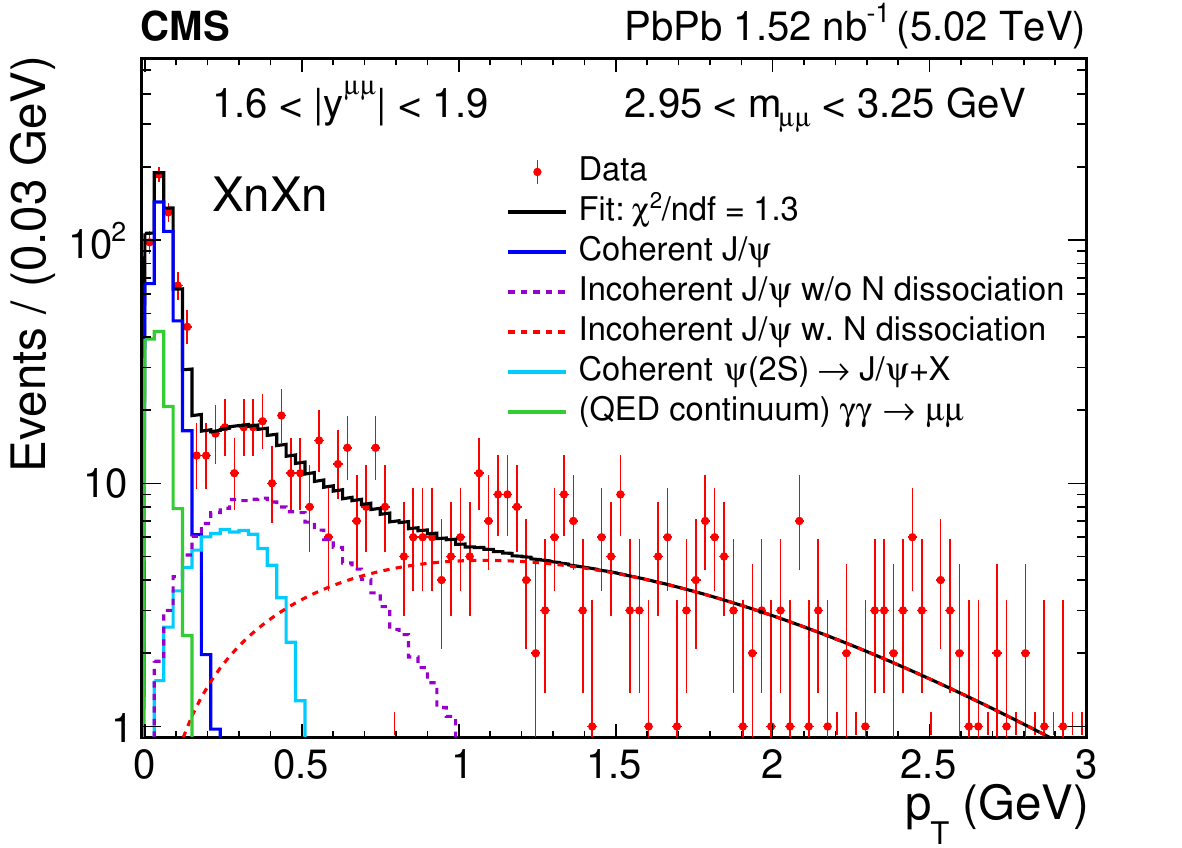}
	\caption{
	The invariant mass distribution (left) of $\mumu$ pairs in the dimuon $\pt<0.2\GeV$ range from the 0n0n (upper) and XnXn (lower) neutron classes. The results of the fit are shown by the various curves, and the corresponding values of the $\chi^2$/ number of degrees of freedom (ndf) and the fitted number of \PJGy and \PGyP{2S} candidates are given. The values of the factors $R_{N}$, $R$, and $f_{\mathrm{D}}$, defined below, are also shown. The transverse momentum distribution (right) of $\mumu$ pairs with a dimuon invariant mass in the $\PJGy$ mass window of ($2.95<m_{\mumu}<3.25\GeV$) in the 0n0n (upper) and XnXn (lower) neutron classes. Again, the results of the fit are shown by the various curves. In both plots, the vertical bars on the data points represent statistical uncertainty.
	}
	\label{fig:SignalsFitting}
\end{figure*}

Examples of invariant mass distribution for the reconstructed dimuon pairs are shown for events with a dimuon $\pt<0.2\GeV$ in Fig.~\ref{fig:SignalsFitting} (left). 
The parameters of the Crystal Ball functions left free in the fit include the pole mass $\PGm_{\PJGy}$ and width $\Gamma_{\PJGy}$ (dominated 
by the experimental momentum resolution) of the $\PJGy$. 
The analogous parameters for the \PGyP{2S} are constrained by demanding $\PGm_{\PGyP{2S}} = \PGm_{\PJGy} M_{\PGyP{2S}}/M_{\PJGy}$ and $\Gamma_{\PGyP{2S}} = \Gamma_{\PJGy} M_{\PGyP{2S}}/M_{\PJGy}$ (as the percentage momentum resolution remains constant within this mass range), respectively, where $M_{\PJGy}$ and $M_{\PGyP{2S}}$ are the world-average mass values~\cite{PDG2022}.
The results of the fit are shown by the various curves, with the $\PJGy$ and $\PGyP{2S}$ signals being clearly visible. The raw yields of $\PJGy$ and $\PGyP{2S}$ events, $N_{\PJGy}$ and $N_{\PGyP{2S}}$, are given in the figure.

The raw number of $\PJGy$ signal events contains contributions from several physics processes: 
coherent and incoherent (with or without nucleon dissociation) $\PJGy$ production from direct photon-nucleus interactions, and also $\PJGy$ mesons coming from the decay of coherent and incoherent $\PGyP{2S}$ production, referred to as ``feed-down''. Coherent and incoherent productions each have distinct
characteristic \pt spectra, so their individual contributions are disentangled using a template fit to the measured $\PJGy$ \pt distribution, as done in Refs.~\cite{Acharya:2019vlb,ALICE:2021gpt}. The feed-down contribution is estimated by measuring the $\PGyP{2S}$ signal yield from the fit to the dimuon invariant mass distributions.
Figure~\ref{fig:SignalsFitting} (right) shows examples of the $\mumu$ \pt spectrum for events with a dimuon invariant mass in the $\PJGy$ mass window ($2.95<m_{\mumu}<3.25\GeV$). Because the different physics processes involved have distinct shapes, their individual contributions can be measured via template fits. The template \pt spectra for coherent $\PJGy$ production, feed-down $\PJGy$ from coherently photoproduced $\PGyP{2S}$ decays, incoherent $\PJGy$ production without nucleon dissociation, and the dimuon continuum from quantum electrodynamic processes ($\PGg\PGg \to \PGmp\PGmm$) are obtained from simulated events with the \STARlight event generator~\cite{Klein:2016yzr} and the $\GEANTfour$~\cite{Agostinelli:2002hh} simulation of the CMS detector. The predicted slope of the \pt distribution for coherently photoproduced $\PJGy$ particles from \STARlight is
adjusted by varying the nuclear radius parameter to best describe the data \pt distribution.

{\tolerance=1200
To describe incoherent $\PJGy$ \pt spectra with nucleon dissociation, an empirical function employed by the H1~\cite{H1:2013okq} and ALICE~\cite{Acharya:2019vlb,ALICE:2021gpt} experiments is used: $\rd{N}/\rd{\pt} \approx \pt [1+(b_{\text{pd}}/n_{\text{pd}}) \pt^{2}]^{-n_{\text{pd}}}$, where $b_{\text{bd}}$ and $n_{\text{pd}}$ are fit parameters. 
Normalizations of the coherent $\PJGy$, incoherent $\PJGy$ without nucleon dissociation, and incoherent $\PJGy$ with nucleon dissociation components are free parameters in the fit. 
The normalization of the dimuon continuum is fixed to the value determined from the invariant mass fitting, as shown in Fig.~\ref{fig:SignalsFitting} (left). Normalizations of coherent and incoherent \PGyP{2S} feed-down to $\PJGy$ are constrained to those of the prompt coherent and incoherent $\PJGy$ components according to the feed-down fractions (\fD) extracted 
from $\PGyP{2S}$ yields in the invariant mass distribution, as described below. The entire procedure is repeated for each neutron multiplicity class.
\par}

The measured \pt spectrum in the range $0.2<\pt<0.5\GeV$ is not well described by the fit, as was also seen in previous ALICE studies~\cite{Acharya:2019vlb,ALICE:2021gpt}. This discrepancy is not unexpected since the \STARlight event generator does not precisely describe all the physics processes involved. To estimate the systematic uncertainty due to this effect, the \pt upper limit is changed to $\pt<0.3\GeV$ and the extraction of the coherent cross section is repeated, leading to an uncertainty of less than 2\%.

The final raw coherent $\PJGy$ yield, $N^{\text{coh}}_{\PJGy}$, is evaluated by:
\begin{equation}
N^{\text{coh}}_{\PJGy}= \frac{N_{\PJGy}}{1+\fI+\fD},
\end{equation}
where \fI and \fD are the fractional contamination
of incoherent processes and coherent $\PGyP{2S}$ feed-down, respectively. 
The incoherent contamination, \fI, is obtained by fitting the \pt distribution and calculating the fraction of incoherent $\PJGy$ events in the $\pt<0.2\GeV$ region.
The feed-down contamination, \fD, is estimated based on the ratio of the number of $\PGyP{2S}$ to $\PJGy$ events, $R_{N} = N_{\PGyP{2S}} / N_{\PJGy}$, found from the fit to the dimuon invariant mass distribution. This parameter can be expressed as
\begin{multline}
R_{N} = \\
\frac{ \sigma_{\PGyP{2S}} \: \mathcal{B}_{\PGyP{2S} \to \PGm\PGm} \: \epsilon_{\PGyP{2S}} }{ 
(\sigma_{\PJGy} \epsilon_{\PJGy} + \sigma_{\PGyP{2S}} \: \mathcal{B}_{\PGyP{2S} \to \PJGy} \: \epsilon_{\PGyP{2S} \to \PJGy}) \: \mathcal{B}_{\PJGy \to \PGm\PGm}}.
\end{multline}

The $\epsilon_{\PJGy}$ and $\epsilon_{\PGyP{2S} \to \PJGy}$ represent reconstruction efficiencies of the prompt $\PJGy$ and $\PJGy$ from $\PGyP{2S}$ decays.
The quantities $\mathcal{B}$s represent branching fractions for each decay model.

The ratio of the cross sections of coherent $\PGyP{2S}$ to coherent $\PJGy$ production, $R=\sigma_{\PGyP{2S}} / \sigma_{\PJGy}$, can then be expressed in terms of $R_{N}$ as
\begin{multline}
	R =\\  \frac{R_{N} \: \mathcal{B}_{\PJGy \to \PGm\PGm} \: \epsilon_{\PJGy}}{\mathcal{B}_{\PGyP{2S} \to \PGm\PGm} \: \epsilon_{\PGyP{2S}} - R_{N} \: \mathcal{B}_{\PGyP{2S} \to \PJGy} \: \epsilon_{\PGyP{2S} \to \PJGy} \: \mathcal{B}_{\PJGy \to \PGm\PGm}}.
\end{multline}
Finally, the fraction of the coherent $\PJGy$ yield that comes from the feed-down of coherently photoproduced $\PGyP{2S}$ is found using the expression
\begin{equation}
\fD = R \: \frac{\epsilon_{\PGyP{2S} \to \PJGy}}{\epsilon_{\PJGy}} \: \mathcal{B}_{\PGyP{2S} \to \PJGy}.
\end{equation}
The world-average values~\cite{PDG2022} for the \PJGy and \PGyP{2S} branching fractions are used in these equations.

The resulting measured values of the incoherent and feed-down fractions and their uncertainties from the fits are given in Tables~\ref{tab:fIFractions} and \ref{tab:fDFractions} for each of the neutron multiplicity class.

\begin{table*}[h!]
	\centering
	\topcaption{The measured values of the incoherent fractions (\fI) and their uncertainties for each of the neutron multiplicity class.}
	\begin{scotch}{ccccc}
	\fI		& 0n0n 		& 0nXn 		& XnXn 		& AnAn\\
	\hline 
	$1.6 < |y| < 1.9$ & $0.005\pm0.001$     & $0.094\pm0.005$  & $0.092\pm0.011$ & $0.034\pm0.002$   \\
	$1.9 < |y| < 2.1$ & $0.007\pm0.001$     & $0.108\pm0.004$  & $0.080\pm0.006$ & $0.038\pm0.002$   \\
	$2.1 < |y| < 2.4$ & $0.004\pm0.001$     & $0.107\pm0.004$  & $0.096\pm0.008$ & $0.033\pm0.004$   \\
	\end{scotch}
	\label{tab:fIFractions}
\end{table*}

\begin{table*}[h!]
	\centering
	\topcaption{The measured values of the feed-down fractions (\fD) and their uncertainties for each of the neutron multiplicity class.}
	\begin{scotch}{ccccc}
	\fD		& 0n0n 		& 0nXn 		& XnXn 		& AnAn\\
	\hline 
	$1.6 < |y| < 1.9$ & $0.039\pm0.004$     & $0.040\pm0.006$  & $0.062\pm0.011$ & $0.040\pm0.003$   \\
	$1.9 < |y| < 2.1$ & $0.021\pm0.004$     & $0.036\pm0.005$  & $0.053\pm0.009$ & $0.027\pm0.003$   \\
	$2.1 < |y| < 2.4$ & $0.025\pm0.005$     & $0.034\pm0.006$  & $0.035\pm0.011$ & $0.028\pm0.004$   \\
	\end{scotch}
	\label{tab:fDFractions}
\end{table*}
\cleardoublepage \section{The CMS Collaboration \label{app:collab}}\begin{sloppypar}\hyphenpenalty=5000\widowpenalty=500\clubpenalty=5000
\cmsinstitute{Yerevan Physics Institute, Yerevan, Armenia}
{\tolerance=6000
A.~Tumasyan\cmsAuthorMark{1}\cmsorcid{0009-0000-0684-6742}
\par}
\cmsinstitute{Institut f\"{u}r Hochenergiephysik, Vienna, Austria}
{\tolerance=6000
W.~Adam\cmsorcid{0000-0001-9099-4341}, J.W.~Andrejkovic, T.~Bergauer\cmsorcid{0000-0002-5786-0293}, S.~Chatterjee\cmsorcid{0000-0003-2660-0349}, K.~Damanakis\cmsorcid{0000-0001-5389-2872}, M.~Dragicevic\cmsorcid{0000-0003-1967-6783}, A.~Escalante~Del~Valle\cmsorcid{0000-0002-9702-6359}, P.S.~Hussain\cmsorcid{0000-0002-4825-5278}, M.~Jeitler\cmsAuthorMark{2}\cmsorcid{0000-0002-5141-9560}, N.~Krammer\cmsorcid{0000-0002-0548-0985}, L.~Lechner\cmsorcid{0000-0002-3065-1141}, D.~Liko\cmsorcid{0000-0002-3380-473X}, I.~Mikulec\cmsorcid{0000-0003-0385-2746}, J.~Schieck\cmsAuthorMark{2}\cmsorcid{0000-0002-1058-8093}, R.~Sch\"{o}fbeck\cmsorcid{0000-0002-2332-8784}, D.~Schwarz\cmsorcid{0000-0002-3821-7331}, M.~Sonawane\cmsorcid{0000-0003-0510-7010}, S.~Templ\cmsorcid{0000-0003-3137-5692}, W.~Waltenberger\cmsorcid{0000-0002-6215-7228}, C.-E.~Wulz\cmsAuthorMark{2}\cmsorcid{0000-0001-9226-5812}
% ...
\par}
\vskip\cmsinstskip
\dag:~Deceased\\
$^{1}$Also at Yerevan State University, Yerevan, Armenia\\
$^{2}$Also at TU Wien, Vienna, Austria\\
% ...
\end{sloppypar}
\end{document}